\documentclass[11pt]{article}
\usepackage[DIV10]{typearea}

\usepackage[latin1]{inputenc}
\usepackage[english]{babel}

\usepackage{amsfonts}
\usepackage{amsmath}    
\usepackage{amssymb}
\usepackage{upgreek}
\usepackage{wasysym}
\usepackage{slashed}
\usepackage{multirow, booktabs}
\usepackage{afterpage}
\usepackage{exscale}

\usepackage[pdftex,colorlinks,pdfpagelabels]{hyperref}
\usepackage[figure,table]{hypcap} 
\hypersetup{
   bookmarksnumbered,
   pdfstartview={FitV},
   pdfpagemode={UseOutlines},
   pdfauthor={Jasper Hasenkamp},
   pdftitle={Dark and visible matter with broken R-parity and the axion multiplet},
   pdfsubject={scientific publication},
   citecolor={blue},
   linkcolor={blue},
   urlcolor={blue},
   filecolor={blue}
} 

\def\a{\alpha}
\def\b{\beta}

\def\g{\gamma}

\def\m{\mu}

\def\r{\rho}

\def\t{\tau}

\def\D{\Delta}

\def\G{\Gamma}

\def\L{\Lambda}
\def\O{\Omega}

\def\Q{\Theta}

\newcommand{\MeV}{\text{ MeV}}
\newcommand{\GeV}{\text{ GeV}}
\newcommand{\TeV}{\text{ TeV}}
\newcommand{\seconds}{\text{ s}}

\newcommand{\axino}{\ensuremath{\widetilde a}}

\newcommand{\order}[1]{\ensuremath{\mathcal{O}\left(#1\right)}}
\newcommand{\mgrav}{m_{3/2}}
\newcommand{\maxino}{m_{\widetilde a}}
\newcommand{\msax}{m_\text{sax}}
\newcommand{\mgluino}{m_{\widetilde g}}

\newcommand{\mplanck}{\ensuremath{M_{\text{P}}}}

\begin{document}
 
\date{\mbox{ }}

\title{ 
{\normalsize     
31st March 2011 \hfill\mbox{}\\}
\vspace{2cm}
\bf
Dark and visible matter with broken R-parity and the axion multiplet
\\[8mm]}
\author{Jasper Hasenkamp and J\"{o}rn Kersten\\[2mm]
{\small\it II.~Institute for Theoretical Physics, University of Hamburg, Germany}\\
{\small\it Jasper.Hasenkamp@desy.de, Joern.Kersten@desy.de}
}

\maketitle

\thispagestyle{empty}

\vspace{1cm}

\begin{abstract}
A small breaking of R-parity reconciles thermal leptogenesis,
gravitino dark matter and primordial nucleosynthesis.
We find that the same breaking relaxes cosmological
bounds on the axion multiplet.
Naturally expected spectra become allowed
and bounds from late particle decays become so weak that they
are superseded by bounds from non-thermal axion production.
In this sense, the strong CP problem serves as an additional
motivation for broken R-parity.
\end{abstract}

\newpage

\section{Introduction}
 A consistent cosmology has to explain the observed matter composition 
of the Universe.
Standard thermal leptogenesis~\cite{Fukugita:1986hr} provides a simple and elegant
explanation for the origin of matter.
However, thermal leptogenesis cannot be reconciled  with an unstable 
gravitino in supergravity~\cite{Falomkin:1984eu,Ellis:1984eq} unless it is very
heavy.
 Fortunately, the gravitino
is one of the best-motivated candidates for dark matter.
If it is the lightest superparticle,
the condition that relic gravitinos do not overclose the Universe
yields an upper bound on the reheating temperature~\cite{Khlopov:1984pf}.
It is remarkable that a reheating temperature of \order{10^{10} \GeV}
can account for the observed matter from leptogenesis and 
at the same time for cold dark matter in the form of thermally
produced relic gravitinos with a mass of \order{100\GeV}.
Interestingly, the strong CP problem can be solved by
the Peccei-Quinn mechanism~\cite{Peccei:1977hh,Peccei:1977ur} in this cosmological setting.
However, the next-to-lightest superparticle (NLSP) becomes
so long-lived that the strongest conflict between 
both notions arises from NLSP decays that spoil
the success of primordial nucleosynthesis~\cite{Moroi:1993mb}.
This and axion physics lead to tighter constraints
on the axion multiplet~\cite{Hasenkamp:2010if}.

It has been shown that in the case of small  R-parity
 breaking thermal leptogenesis,
gravitino dark matter and primordial nucleosynthesis are
naturally consistent~\cite{Buchmuller:2007ui}.
Since a consistent cosmology should also enable a solution
to the strong CP problem, we study in this letter
how far the restrictions on the axion
multiplet are softened, if R-parity is broken. 

The small R-parity breaking couplings allow the lightest
ordinary superparticle (LOSP) to decay into pairs of
Standard Model particles. These decays happen
instantaneously compared to the Hubble time after
LOSP freeze-out or the R-parity-conserving decay into the gravitino.
 Thus they do not endanger the success
of Big Bang nucleosynthesis (BBN).
In this way the aforementioned NLSP decay problem
is circumvented, which at the same time relaxes the constraint on the 
axino lifetime and thus the allowed range of the axino mass and possibly
of other parameters.
It is allowed to decay right before BBN,
instead of the requirement with conserved R-parity  
to decay before NLSP freeze-out.
Thus its decay temperature can be lowered
by three orders of magnitude.
In addition, saxion decays are allowed to produce
NLSPs at any time before BBN.

The effect on the LOSP decay is the crucial
impact of broken R-parity on the constraints 
on the axion multiplet. 
All members of the axion
multiplet (axion a, axino $\widetilde a$ and saxion $\phi_\text{sax}$)
obtain additional R-parity breaking couplings, 
but
R-parity breaking interactions and decays are suppressed
by the Peccei-Quinn scale and additionally by
the small R-parity breaking couplings.
Thus they are produced and decay in the same way as
if R-parity were conserved.
One exception to this statement occurs, if
the axino is the NLSP\@.
In the following
we fix the gravitino mass to be of \order{100 \GeV} 
and the reheating temperature
at a rather large value of \order{10^{10} \GeV}.
This is a worst case scenario for the axion multiplet.
We will comment on
the DFSZ~\cite{Dine:1981rt,Zhitnitsky:1980tq} and KSVZ~\cite{Kim:1979if,Shifman:1979if} invisible axion models,
especially regarding the possibility of
an axino next-to-NLSP\@. 
Nevertheless our analysis
is general and can be applied to 
axions in superstring models~\cite{Witten:1984dg}.
Our main results are comprised in Table~\ref{tab:1}.
Since, for example, naturally expected  spectra become allowed,
our results serve as an additional motivation for broken R-parity.

\section{Axino and saxion with conserved R-parity}
\paragraph{Axino}
Even if axinos do not enter thermal equilibrium after
inflation, they are, nevertheless, regenerated by
thermal scatterings and decays in the thermal plasma.
The resulting density can be estimated in units of
today's critical density as~\cite{Covi:2001nw,Brandenburg:2004du,Strumia:2010aa}
\begin{equation}
\label{eq:oaxino}
 \O_{\widetilde a} h^2 \simeq 7.8 \times 10^6 
 \left(\frac{m_{\widetilde a}}{1 \TeV}\right) \left(\frac{T_\text{R}}{10^9 \GeV}\right)   
 \left(\frac{10^{12} \GeV}{f_a}\right)^2 ,
\end{equation}
where $\maxino$ and $T_\text{R}$ denote 
the axino mass and the reheating temperature, respectively.
The axion decay constant $f_a=f_\text{PQ}/N$ with $f_\text{PQ}$ denoting
the Peccei-Quinn scale if $N$ denotes the number of different vacua.  For
the KSVZ (DFSZ) model $N=1 \, (6)$.
The produced LOSP\footnote{The lightest  ordinary superparticle (LOSP) is the
lightest superparticle of the minimal supersymmetric standard model (MSSM).
In this sense, gravitino and axino are extraordinary superparticles.}
 and/or gravitino abundances from axino decay are orders of
magnitude larger than the thermal abundances even if re-annihilation
is taken into account~\cite{Choi:2008zq,Baer:2011hx}.
The produced LOSP abundance is inconsistent with the scenario,
 because the NLSP decay problem is worsened.
 Therefore, the decay is demanded to happen before
LOSP freeze-out, so that one does not have to worry about the produced number of
LOSPs since they thermalise normally. In this case the lower bound on the axino mass reads
\begin{equation}
\label{eq:maxinomin2}
 m_{\widetilde a} \gtrsim 6 \times 10^2 \GeV
 \left( 1 -  \frac{m_{\widetilde g}^2}{\maxino^2}\right)^{-1}
 \left(\frac{m_\text{losp}}{10^2 \GeV}\right)^{\frac{2}{3}} 
 \left(\frac{f_a}{10^{10} \GeV}\right)^{\frac{2}{3}}
   \left(\frac{g_\ast(T^\text{dec}_{\widetilde a})}{100}\right)^\frac{1}{6} ,
\end{equation}
where $\mgluino$ denotes the gluino mass, $g_\ast$ the effective number
of relativistic degrees of freedom of the Universe and 
we have used that the freeze-out temperature $T^\text{fo}_\text{losp} \approx m_\text{losp}/25$.
By $T^\text{dec}_{\widetilde a}$ we denote the temperature of the Universe after the axino decay.
Since in supergravity the axino mass is---as the other superparticle
masses---generically
of the order of the gravitino mass~\cite{Chun:1995hc}, i.e.,
$\maxino \sim m_\text{susy} \sim \mgrav$, the mass bound~\eqref{eq:maxinomin2}
can be interpreted as an upper bound on the axion decay constant,~i.e., $f_a \lesssim 10^{10}\GeV$.
In~\eqref{eq:maxinomin2} the decay into a gluino-gluon pair with width~\cite{Covi:2001nw}
\begin{equation}
 \G_{\axino}^{\widetilde g g} = \frac{\alpha_s^2}{16 \pi^3} 
 \frac{m_{\widetilde a}^3}{f_a^2} \left(  1 - \frac{m_{\widetilde g}^2}{\maxino^2} \right)^3 
 \label{axinogluinogluon}
\end{equation}
is assumed to be dominant and the strong coupling constant $\a_\text{s}(\m)=\a_\text{s}(\maxino)\simeq 0.1$.
The involved operator exists for any axion model that is able to solve the strong CP problem.
Investigating other axino decay channels in the KSVZ model, we find
that they require $\maxino$
to be larger than the expected gluino mass, which
keeps~\eqref{axinogluinogluon} to be the dominant decay channel.
Independently of $f_a$ the axino must be sufficiently heavier than the gluino, which 
is expected to be among the heavier superparticles
due to the running of its mass. For instance, with
a gluino mass $\mgluino = 1 \TeV$ and the parameter values appearing in~\eqref{eq:maxinomin2}
 the axino mass is required to be larger
than about $1.35 \TeV$.

The DFSZ superfield is directly coupled to the Higgs superfields. Therefore, the DFSZ axino
may dominantly decay into a Higgsino-Higgs pair. Using
$\mathcal{L}= \frac{2 \mu}{f_a} (H_u^0 \widetilde H_d^0 + H_d^0 \widetilde H_u^0 ) \,\axino + \text{h.c.}$,
we obtain in the decoupling limit\footnote{The result seems quite insensitive to the possibility of the Guidice-Masiero mechanism
to generate the $\mu$-term, but this might be different for the saxion decay into two light Higgses~\cite{Jeong:2011xu}. For the neglected
phase space factor we find
$
\lbrace (1+x_\chi - x_h)^3 (1+ x_\chi + x_h)^3 ((1-x_\chi)^2 - x_h^2) \rbrace^{1/2}
$ with $x_\chi= m_{\chi_i^0}/\maxino$ and $x_h =m_h/\maxino$.
}
\begin{equation} \label{axinohiggsinohiggs}
 \G (\axino \rightarrow \widetilde H + h) \simeq
 \frac{\maxino}{4 \pi} \, \sum_i |N_{\chi_i^0 \widetilde H_d^0} \sin{\beta} + N_{\chi_i^0 \widetilde H_u^0} \cos{\beta}|^2  
 \left(\frac{\mu}{f_a} \right)^2 \, ,
\end{equation}  
where $\mu$ is the usual supersymmetric $\mu$-parameter, 
$\tan{\b}=v_u/v_d$ is
the ratio of the Higgs vacuum expectation values and
$N_{\chi_i^0 \widetilde H_d^0}$ and $N_{\chi_i^0 \widetilde H_u^0}$ denote the Higgsino fractions
of the $i$-th neutralino.
We consider the case of degenerate, maximally mixed Higgsinos, which are then decoupled from the gauginos.
For sufficiently large $\tan{\beta}$ the mixing factor
in~\eqref{axinohiggsinohiggs} then becomes about $1/2$.
Thus, the sum of both Higgsinos softens the mass bound to
\begin{equation}
\label{maxinomindfsz}
 \maxino^\text{dfsz} \gtrsim 5.4 \times 10^2 \GeV 
 \left( \frac{f_a}{10^{11} \GeV} \right)^2
 \left( \frac{100 \GeV}{\mu} \right)^2
 \left( \frac{m_\text{losp}}{10^2 \GeV} \right)^2
 \left( \frac{g_\ast(T^\text{dec}_{\axino})} {100} \right)^\frac{1}{2} .
\end{equation} 
The upper bound on the Peccei-Quinn scale from late axino decay
may thus be softened to $f_a \lesssim 10^{11} \GeV$
and for $f_a \lesssim 10^{10} \GeV$ we observe that the open decay channel into
Higgsino-Higgs suffices in the DFSZ model. Other decay
channels are subdominant compared to \eqref{axinogluinogluon} and
\eqref{axinohiggsinohiggs}.

Altogether, only spectra with $\maxino > \mgluino$ are allowed in all axion models.
This can be viewed as a problem, because it requires the axino to be heavier than naturally expected.
The bound may become softened to $\maxino > m_{\widetilde H} + m_h$ in the DFSZ model for a small $\mu$-parameter.

\paragraph{Saxion}
Due to supersymmetry the couplings of the saxion have the same strength as the
axino couplings. Thus the saxion is produced as efficiently as the axino.
Its dominant decay producing MSSM particles is into a pair of gluons with~\cite{Baer:2010gr} 
\begin{equation}
\label{saxgg}
 \G_\text{sax}^{gg} = \frac{\alpha_s^2}{32 \pi^3} 
 \frac{m_\text{sax}^3}{f_a^2} 
\end{equation}
or---for the DFSZ model---into a pair of light Higgses with~\cite{Kawasaki:2011ym,Jeong:2011xu}
\begin{equation}
 \G(\phi_\text{sax} \rightarrow 2 h) = \frac{\msax}{8 \pi}
 \left( 1 - \frac{4 m_h^2}{\msax^2} \right)^\frac{1}{2}
 \left( \frac{\mu}{f_a} \right)^2
 \left( \frac{\mu}{\msax} \right)^2 \, . 
\end{equation} 
Since in these
decays no superparticles are produced, they may happen right before BBN\@. 
Since the saxion receives
its mass from SUSY breaking, one expects $m_\text{sax} \sim m_\text{susy}$.
If the saxion is heavy enough to produce superparticle pairs,
its decay could lead to the same worsening of the NLSP decay problem
as the axino decay.
In  this situation the lower bound on the saxion mass from the decay
into a pair of gluons becomes
in the end 
\begin{equation}
\msax \gtrsim 7.6 \times 10^2 \GeV 
\left(\frac{m_\text{losp}}{10^2 \GeV}\right)^\frac{2}{3}
\left(\frac{f_a}{10^{10} \GeV}\right)^\frac{2}{3}
\left(\frac{g_\ast(T_\text{sax}^\text{dec})}{100}\right)^\frac{1}{6} \, .
\end{equation} 
If the DFSZ saxion decays dominantly into Higgs pairs, we can derive
an upper bound on the saxion mass
\begin{equation}
 \msax \lesssim 740 \GeV
 \left( 1 - \frac{4 m_h^2}{\msax^2} \right)^\frac{1}{2}
 \left( \frac{10^{11} \GeV}{f_a} \right)^2
 \left( \frac{\mu}{300 \GeV} \right)^4
 \left( \frac{10^2 \GeV}{m_\text{nlsp}} \right)^2
 \left( \frac{100}{g_\ast(T^\text{dec}_\text{sax})} \right)^\frac{1}{2}
\end{equation} 
We see that for $f_a \lesssim 10^{10} \GeV$ the DFSZ saxion mass is
only required to allow for the decay into a pair of light Higgses.

In addition, the saxion may decay as well into 
two axions with~\cite{Kawasaki:2007mk}
\begin{equation} \label{eq:saxtoaa}
 \G^{aa}_\text{sax} \simeq \frac{x^2}{64 \pi} \frac{m_\text{sax}^3}{f_a^2} \, ,
\end{equation}
where the self-coupling $x$ can be of order $1$. 
The produced axions represent a form of dark radiation, i.e.,
decoupled, relativistic particles not present in the Standard Model.
During BBN the energy density of
dark radiation $\r_\text{dr}$ is constrained to be less than the 
energy density of one additional neutrino species~\cite{Pospelov:2010hj}, which
translates into
\begin{equation}
\label{boundonrhodr}
 \left.\frac{\r_\text{dr}}{\r_\text{SM}}\right|_\text{BBN} \lesssim 0.14 \, .
\end{equation}
Here $\r_\text{SM}$ denotes the energy density as expected from
the Standard Model
and
\begin{equation}
\r_\text{dr}|_\text{BBN}= B_{aa}\, \r_\text{sax}|_{T^\text{dec}_\text{sax}} 
\left(\frac{g_{\ast s}(T_\text{BBN})}{g_\ast(T^\text{dec}_\text{sax})}\right)^{4/3}
 \left(\frac{T_\text{BBN}}{T^\text{dec}_\text{sax}}\right)^4 ,
\end{equation}
where $g_{\ast s}$ counts the relativistic degrees of freedom
determining the entropy density $s$ and
$g_{\ast s}(T^\text{dec}_\text{sax}) = g_\ast(T^\text{dec}_\text{sax})$.
Written as bound on the branching ratio $B_{aa}$
of the saxion into two axions~\eqref{boundonrhodr} reads\footnote{
Our expression differs from
the literature in the numerical value and the dependence on $g_\ast$. 
This is because we have taken into account properly the scaling of $\r_\text{dr} \propto g_{\ast s}^{4/3} T^4$.
The simplifying assumption of entropy conservation $(g_{\ast s} = const.)$,
 which is usually made, does indeed not hold during these stages of the Universe.
} 
\begin{equation}
\label{boundonbaa}
 B_{aa} \lesssim 0.4 \left(1+ 50 \pi^2  x^2 \right)^\frac{1}{2}
  \left(\frac{10^{10} \GeV}{f_a}\right) \left(\frac{\msax}{10^2 \GeV}\right)^\frac{1}{2} 
 \left(\frac{Y_\text{sax}^\text{eq}}{Y_\text{sax}}\right) \left(\frac{g_\ast(T_\text{sax}^\text{dec})}{10.75}\right)^\frac{1}{12} \,.
\end{equation}
In this inequality we have approximated the decay width of the
saxion as
$\G_\text{sax} \simeq \G_\text{sax}^{gg} + \G_\text{sax}^{aa}$. 
Thus our conclusion should hold qualitatively for any axion model.
Then the branching ratio reduces to
\begin{equation}
 B_{aa} \simeq \frac{\G_\text{sax}^{aa}}{\G_\text{sax}^{aa}+\G_\text{sax}^{gg}} = \frac{x^2}{x^2+2 \a_s^2/\pi^2} \, .
\end{equation}
The value for $f_a$ appearing in~\eqref{boundonbaa} corresponds to the upper bound
on $f_a$ from axino decay and the axion energy density $\O_a$ (see below). Since in our scenario the reheating temperature
is fixed at rather large values, all members of the axion multiplet enter thermal equilibrium after inflation
for such small values of $f_a$. Therefore the saxion yield $Y_\text{sax}$ cannot be smaller than the
equilibrium value $Y_\text{sax}^\text{eq} \simeq 1.21 \times 10^{-3}$.
The appearing values for $f_a$ and $\msax$  are chosen to
show the worst situation in the considered scenario.
Like a smaller $f_a$ or a larger $\msax$,
also a larger $x$ leads to
an earlier decay, which corresponds to a smaller $\O_\text{sax}$
at its decay. Thus there is a self-curing effect for large $x$.
 The bound~\eqref{boundonbaa} represents indeed an implicit equation for the
 self-coupling $x$. Evaluating it for $x$ it turns out that there is
 no constraint on $x$ at all in the scenario under consideration.

We point out that the absence of any 
bound on $x$ is due to the expectation $\msax\sim m_\text{susy}$
and the restriction of $f_a$ to small values appropriate
for the considered scenario.
Furthermore, $Y_\text{sax}$ could be much larger
than $Y_\text{sax}^\text{eq}$,
if the saxion is produced from coherent 
oscillations after inflation
as discussed below.

The saxion might decay before LOSP freeze-out, if the self-coupling is
strong enough. Neglecting conservatively the saxion decay into gluons and Higgses
we find
\begin{equation}
 x \gtrsim 0.9 
 \left(\frac{10^2 \GeV}{\msax}\right)^\frac{3}{2}
 \left(\frac{f_a}{10^{10} \GeV}\right)
 \left(\frac{m_\text{losp}}{10^2 \GeV}\right)
 \left(\frac{g_\ast(T_\text{sax}^\text{dec})}{61.75}\right)^\frac{1}{4} \, .
\end{equation} 
In this situation there is no additional constraint on the saxion mass.
Note that the decay into axions does not produce a significant
amount of entropy.

If a new best-fit value demands additional radiation energy in the Universe~\cite{Aver:2010wq,Izotov:2010ca,Dunkley:2010ge,Hamann:2007pi}, which
is often parameterised by a change of the effective neutrino degrees of freedom $\D N_\text{eff} > 0$,
we can determine parameter
values from~\eqref{boundonbaa}, such that
the additional energy is formed by axions from
saxion decay.
For $f_a \leq 10^{10} \GeV$ and $\msax \geq 10^2 \GeV$ the maximal $\D N_\text{eff}$ is $0.6$.
However, when these requirements are relaxed also larger $\D N_\text{eff}$ are possible.
For example, we obtain $\D N_\text{eff} \simeq 1$ with a rather small saxion mass $\msax =10 \GeV$,
 $f_a=10^{10} \GeV$ and a self-coupling $x =0.1$.

\section{R-parity violating case}
\paragraph{Axino}
Since in the case of broken R-parity the NLSP decay problem is absent,
the axino may decay right before BBN\@. Produced superparticles decay promptly
into particles of the Standard Model, which thermalise normally. Then the lower bound~\eqref{eq:maxinomin2} becomes
\begin{equation}
\label{eq:maxinomin3}
 m_{\widetilde a} \gtrsim 89 \GeV 
 \left( 1- \frac{\mgluino^2}{\maxino^2}\right)^{-1}
 \left(\frac{f_a}{10^{12} \GeV}\right)^{\frac{2}{3}}
 \left(\frac{T^\text{dec}_\text{min}}{4 \MeV}\right)^\frac{2}{3} ,
\end{equation}
where we still assume the decay channel
$\widetilde a \to \widetilde g g$ to dominate 
and take $T^\text{dec}_\text{min}$ as lower bound on the temperature
after the particle decay.
The bound $T^\text{dec}_\text{min} = 4 \MeV$,
corresponding to a lifetime $\t_\text{max} \sim 0.05 \seconds$,
considers the neutrino energy density during the 
neutrino thermalisation
process~\cite{Kawasaki:1999na,Adhya:2003tr,Hannestad:2004px}, if
 the particle dominated the energy density of the Universe before its decay.
This is likely to happen in the considered scenario (see below).
Weaker bounds $T^\text{dec}_\text{min} \sim 0.7 \MeV$ arise from BBN calculations~\cite{Kawasaki:2000en,Ichikawa:2005vw}.
As in the following we omit the dependence on $g_\ast(T^\text{dec}_{\widetilde a}=T^\text{dec}_\text{min})=10.75$.
We see that due to $T^\text{dec}_\text{min} \ll T^\text{fo}_\text{nlsp}$
 the axino mass bound becomes so weak that masses much smaller
than the expected gluino mass would become allowed. 

In this situation
 the axino
decays always depend on which channels are kinematically open and thus, in principle, on the full spectrum.
This is a qualitative difference to the R-parity conserving case.
Particularly interesting is the possibility of an axino next-to-NLSP,
which we will assume in the following.
Then its dominant decay is
fixed into the NLSP, in other words, into the LOSP\@.
From~\eqref{eq:maxinomin3} we see that the NLSP were allowed to be a gluino.
A recent analysis reports $\mgluino > 322 \GeV$ as experimental
lower limit for such a long-lived gluino~\cite{Farrar:2010ps}.
From~\eqref{maxinomindfsz} we see that the DFSZ model would in addition allow for
a Higgsino NLSP, if the corresponding decay channel were kinematically open.

Since the lightest neutralino is likely one of the lightest superparticles in the spectrum,
 one interesting decay is into neutralino and photon with~\cite{Covi:2001nw}
\begin{equation} \label{eq:GammaAxinoNeutralino}
 \G(\widetilde a \rightarrow \chi^0_i + \g)=
 \frac{\a^2_\text{em}C^2_{a \chi^0_i \g}}{128 \pi^3} \frac{m_{\widetilde a}^3}{f_a^2} 
 \left( 1-\frac{m^2_{\chi^0_i}}{\maxino^2} \right)^3 ,
\end{equation}
where $C_{a \chi^0_i \g}= ( C_{aBB}/ \cos{\Q_W}) N_{\chi^0_i \widetilde B^0}$, while
$N_{\chi^0_i \widetilde B^0}$ is the bino fraction of the $i$-th neutralino and
$\Q_W$ denotes the weak mixing angle. 
We take the electromagnetic coupling constant $\a_\text{em}(\m)=\a_\text{em}(\maxino)\simeq 1/128$.
The axion to two $B$ bosons coupling $C_{aBB}$ varies for different implementations of
different axion models.%
\footnote{For example, in the DFSZ model with $(d^c,\,e)$ unification, $C_{aBB}=8/3$.
In the KSVZ model, for different electromagnetic charges of the heavy quark
$e_\text{Q}=0,\,-1/3,\,2/3$, $C_{aBB}=0,\,2/3,\,8/3$, respectively.
Below the QCD scale, $C_{aBB}$ is reduced by 1.92~\cite{Kim:1998va}.}
For simplicity we set
$C_{aBB}=N_{\chi^0_1 \widetilde B^0}=1$. We find 
\begin{equation}
 \label{eq:maxinomin4}
 m_{\widetilde a} \gtrsim 45 \GeV 
 \left( 1- \frac{m_{\chi^0_1}^2}{\maxino^2} \right)^{-1}
 \left(\frac{f_a}{10^{10} \GeV}\right)^{\frac{2}{3}}
 \left(\frac{T^\text{dec}_\text{min}}{4 \MeV}\right)^\frac{2}{3}
\end{equation}
and conclude that a neutralino with a substantial bino component
can be the NLSP, if the axino is the next-to-NLSP\@.
Thus a superparticle spectrum with $m_{\widetilde g}> \maxino > m_{\chi_1^0}$
is now easily possible without any BBN conflict.

If the axino decay is fixed into a sneutrino, it decays via an 
intermediate neutralino into a photon and a sneutrino-neutrino pair,
i.e., $\axino \rightarrow \g \chi^{0\ast} \rightarrow \g \widetilde \nu \nu$.
This process is suppressed compared to~\eqref{eq:GammaAxinoNeutralino}
by an additional power of $\a_\text{em}$ and further factors
depending on the neutralino composition and the exact spectrum.
Thus in case of a sneutrino NLSP an axino next-to-NLSP might be possible only for
parameter values at the boundaries of the allowed region and a tuned spectrum.
Consequently, this situation is disfavoured.

For other sfermion NLSPs the situation depends much more strongly on the axion model.
In the DFSZ model we estimate the kinematically unsuppressed decay width of the axino into
a stop-top pair using the Lagrangian of~\cite{Nieves:1986ed} as
\begin{equation}
\label{eq:att}
 \G (\axino \rightarrow \widetilde t + t) \simeq \frac{\maxino}{16 \pi} \left( \frac{m_t X_u}{f_a} \right)^2 ,
\end{equation}
where $m_t$ denotes the fermion mass and $X_u=1/(\tan^2\b + 1)$.
This decay arises from a dimension-four operator that becomes important
at low decay temperatures and accordingly small masses.
If only this channel were open, the resulting
lower mass bound on the axino would become
\begin{equation}
\label{eq:maxinomin5}
 \maxino \gtrsim  2.4 \times 10^2 \GeV
 \left(\frac{f_a}{10^{12} \GeV}\right)^2
 \left(\frac{T^\text{dec}_\text{min}}{4 \MeV}\right)^2
   \left(\frac{\tan\beta}{10}\right)^4
  \left(\frac{173\GeV}{m_t}\right)^2 .
\end{equation}
The lower bound would practically be given by $f_a$ and the requirement to have
the channel kinematically unsuppressed, since a long-lived
stop with mass below $249 \GeV$ is excluded according to the CDF
experiment~\cite{Aaltonen:2009kea}.
It might well happen that the decay of the next-to-NLSP (\axino) into
the NLSP ($\widetilde t$\,) and its superpartner (t) is kinematically forbidden.
The axino would decay violating R-parity. This is the same if the
\axino~is the NLSP\@.
We comment on this excluded case below.

The decay width into other sfermions is given by~\eqref{eq:att}, if
$m_t$ is replaced by the corresponding fermion mass and $X_u \rightarrow X_d=\tan^2\b/(1+\tan^2\b)$
if appropriate.
For example, the decay width into stau and tau is~\eqref{eq:att} with the
replacements $m_t \rightarrow m_\tau \simeq 1.78 \GeV$ and $X_u \rightarrow X_d$.
The corresponding lower bound on the axino mass becomes
\begin{equation}
\label{eq:maxinomin6}
 \maxino \gtrsim 2.3 \times 10^2 \GeV
 \left(\frac{f_a}{10^{12} \GeV}\right)^2
 \left(\frac{T^\text{dec}_\text{min}}{4 \MeV}\right)^2 .
\end{equation}
Thus, for $f_a$ below $10^{12} \GeV$ also this channel is restricted
by the requirement of kinematic accessibility only. Since
$m_\tau/m_{\widetilde \tau} \ll 1$, this is not a strong
constraint, if the stau is not much heavier than the gravitino.
Consequently, if the stau is the NLSP, the DFSZ axino may be the next-to-NLSP.

Since the other leptons are lighter, in these cases the bound
becomes tighter. While the smuon stays possible, the selectron
would require $\maxino \gtrsim 1 \TeV$ even for parameter
values at the boundaries.
These considerations expand to the quarks lighter than the top.
The superpartners of the bottom, charm, and strange quarks are
possibilities.  For the down squark the Peccei-Quinn scale needs to be
close to the lower limit.
The situation for the up squark is even worse than the one for the
selectron.

In the KSVZ model axino-sfermion-fermion interactions are one-loop-suppressed
in the low-energy effective theory~\cite{Covi:2002vw,Covi:2004rb}, which weakens axino to
sfermion-fermion decays substantially relative to the DFSZ case.
The tree-level decay via intermediate neutralino into other
sfermion-fermion pairs is comparable to the decay
into sneutrino-neutrino.
Consequently, the KSVZ axino is disfavoured, if the axino decay is fixed
into a sfermion-fermion pair. 

If the axino were the NLSP, i.e., for a spectrum
with $m_\text{losp} > \maxino > \mgrav$, it would decay either into the
gravitino or via R-parity violation.  Both decay modes are strongly
suppressed by the Planck scale or by the Peccei-Quinn scale and the
R-parity violating coupling, respectively.  Therefore, the lifetime of
the axino would always become much larger than the time of
BBN\@. This is independent of whether R-parity is broken by bilinear or
trilinear couplings~\cite{Hooper:2004qf,Chun:2006ss,Kim:2001sh}.
Since $\O_{\axino} \gg 1$, such a late decay would spoil the predictions
of BBN\@.
We conclude that an axino NLSP stays excluded
in the R-parity violating scenario.

\paragraph{Saxion}
If R-parity is violated also
superparticle pairs from saxion decays are harmless. If its dominant 
decay channel is into a pair of (massless) gluons, the most
severe bound on the saxion mass is found to be
similar to~\eqref{eq:maxinomin3}.
So for $f_a \sim 10^{10} \GeV$ the saxion mass is practically
not constrained in the R-parity violating case.
However, considerations concerning the self-coupling $x$ are not affected
by R-parity violation.

If late-decaying particles enter thermal equilibrium
after inflation, they likely dominate the energy density
of the Universe at their decay, since the energy density of matter $\r_\text{mat}$
grows relative to the energy density of radiation $\r_\text{rad}$
as the scale factor $a$ of the Universe,~i.e., $\r_\text{mat}/\r_\text{rad} \propto a$.
If a particle dominates the energy density of the Universe,
 it might produce considerable entropy at its decay, which
 could dilute cosmological abundances to too small values.
Axino and saxion enter thermal equilibrium at
$T_\text{R} \sim 10^{10} \GeV$ if $f_a \lesssim 10^{11.5} \GeV$.
The dilution factor due to their decay can be estimated to be~\cite{Lyth:1993zw,Hasenkamp:2010if}
\begin{equation}
 \D \simeq 2 
 \left(\frac{f_a}{10^{11} \GeV}\right) 
 \left(\frac{1 \TeV}{m_{\widetilde a /\text{sax}}}\right)^\frac{1}{2} 
 \left(\frac{g_\ast(T^\text{dec})}{61.75}\right)^\frac{1}{4} ,
\end{equation}
assuming the decay into a gluino-gluon pair~\eqref{axinogluinogluon}
 and two gluons~\eqref{saxgg}, respectively, to be dominant.
The numerical difference between axino and saxion is small.
Here, the upper bound on the dilution factor $\D \geq 1$
arises from leptogenesis itself and is $\D^\text{max} \sim 10^4$.
Thus the occurring dilution is so small that there arises no constraint
from entropy production
by the thermally produced axion multiplet.

The saxion may also be produced by coherent oscillations around its potential minimum
after inflation. The temperature at the onset of saxion oscillations is
\begin{equation}
 T_\text{sax}^\text{osc} \simeq 2.2 \times10^{10} \GeV 
 \left(\frac{m_\text{sax}}{1\TeV}\right)^\frac{1}{2}
  \left(\frac{228.75}{g_\ast(T_\text{sax}^\text{osc})}\right)^\frac{1}{4} .
  \label{Toscsax}
\end{equation}
If the reheating temperature $T_\text{R} < T_\text{sax}^\text{osc}$ as in the
standard scenario,
the produced saxion abundance is~\cite{Kawasaki:2007mk}
\begin{align}
 \frac{\rho^\text{osc}_\text{sax}}{s} &=
 \frac{1}{8} T_\text{R} \left(\frac{\phi_\text{sax}^\text{i}}{M_\text{Pl}}\right)^2
\nonumber\\
 &\simeq 4.2 \times 10^{-9} \GeV \left(\frac{T_\text{R}}{2 \times 10^9\GeV}\right)
 \left(\frac{f_a}{10^{10}\GeV}\right)^2 
 \left(\frac{\phi_\text{sax}^\text{i}}{f_a}\right)^2 ,
 \label{eq:rhoosc}
\end{align}
where $\phi_\text{sax}^\text{i}$ denotes the initial amplitude of the oscillations
and where we have assumed the simplest saxion potential,
$V = \frac{1}{2} m_\text{sax}^2 \phi_\text{sax}^2$.

The dilution factor due to saxion decays $\D \simeq 0.75 \, T^=_\text{sax}/T^\text{dec}_\text{sax}$
with $T^=_\text{sax} = \frac{4}{3} \frac{\rho^\text{osc}_\text{sax}}{s}$
is given by
\begin{equation}
 \Delta \simeq 4 \times 10^{-10}
 \left(\frac{T_\text{R}}{2 \times 10^9\GeV}\right) 
 \left(\frac{f_a}{10^{10}\GeV}\right)^3 
 \left(\frac{\phi_\text{sax}^\text{i}}{f_a}\right)^2
 \left(\frac{1\TeV}{m_\text{sax}}\right)^\frac{3}{2}
 \left(\frac{g_\ast(T^\text{dec}_\text{sax})}{10.75}\right)^\frac{1}{4} .
\end{equation}
Since the dilution factor is by definition $\geq 1$, values for $\D$ smaller
than one lead to the standard scenario with $\D=1$. 
In this discussion we fix the reheating temperature at its lower
boundary from thermal leptogenesis, $T^\text{min}_\text{R}=2 \times 10^9 \GeV$~\cite{Davidson:2002qv}
or $2 \Delta \times 10^9 \GeV $ with $\Delta>1$.
The tightest bound on the initial amplitude in the standard scenario is found for
the maximal axion decay constant $f_a = 10^{10}\GeV$ and
a small saxion mass, for concreteness $\msax = 10^2 \GeV$.
It is $\phi^\text{i}_\text{sax} \lesssim  7 \times 10^{13} \GeV$.
In comparison, the loosest bound is found for the minimal axion decay constant
$f_a = 6 \times 10^8 \GeV$ and a rather large saxion mass, for
concreteness $\msax = 1 \TeV$.
It is $\phi^\text{i}_\text{sax} \lesssim 1.4 \times 10^{15} \GeV$.
Thus both bounds are far below the Planck scale. 
Bounds on the initial amplitude of saxion oscillations are not affected
by R-parity violation. Conservatively allowing for smaller saxion
masses as well, we summarise these bounds in Table~\ref{tab:1} as
$\phi^\text{i}_\text{sax} \lesssim ( 10^{13} \text{--} 10^{15}) \GeV$.

In the case of considerable entropy production, for a certain $\Delta$
the reheating temperature has to become larger than the temperature
at the onset of saxion oscillations, i.e., $T_\text{R} >T_\text{sax}^\text{osc}$.
Then $T_\text{R}$ in~\eqref{eq:rhoosc} has to be replaced by $T_\text{sax}^\text{osc}$ from~\eqref{Toscsax}.
In this case larger initial amplitudes may be allowed, while at the same
time larger axion decay constants become allowed.
The tightest bound on the initial amplitude in the scenario with $\D=\D^\text{max}=10^4$
is found for the maximal axion decay constant $f_a=4 \times 10^{12} \GeV$ and
the minimal saxion mass $\msax \sim 300 \GeV$ from the decay into a gluon pair. 
It is $\phi^\text{i}_\text{sax} \lesssim 4 \times 10^{14} \GeV$.
In comparison, the loosest bound is found for the minimal axion decay constant
$f_a = 6 \times 10^8 \GeV$ and a rather large saxion mass, for
concreteness $\msax= 1 \TeV$.
It is $\phi^\text{i}_\text{sax} \lesssim 5 \times 10^{16} \GeV$.
We summarise these bounds, that are still far below the Planck scale,
 in Table~\ref{tab:1} as a range
$\phi^\text{i}_\text{sax} \lesssim  5 \times (10^{14} \text{--}  10^{16}) \GeV$.

\section{Constraints from the axion}
First, the lower limit
on the axion decay constant from axion physics~\cite{Raffelt:2006cw},
$f_a \gtrsim 6 \times 10^8 \GeV$,
is not changed by the additional R-parity violating
interactions.
Furthermore, the axion still decays harmlessly into two photons with a lifetime much
larger than the age of the Universe.
Due to its tiny mass, the thermal relic density of axions
is negligible. 
In contrast, a large axion density may be produced non-thermally by
vacuum misalignment and topological defects. 

The density produced by 
vacuum misalignment~\cite{Preskill:1982cy,Abbott:1982af,Dine:1982ah} is usually given as
 \begin{equation}
 \label{eq:oa1}
  \O_a^\text{mis} h^2 \sim a_0^2 \left(\frac{N \,f_a}{10^{12} \GeV }\right)^{7/6} ,
 \end{equation}
where $a_0$ comprises model-dependent factors.
In the presence of gravitino dark matter we require $\O_a/\O_\text{DM}=r \ll 1 $,
which gives an upper bound 
\begin{equation}
\label{upperfaboundstan}
f_a < 10^{10} \GeV \quad \text{for} \quad a_0 = N = 1 \;,\; r=0.04 \, . 
\end{equation}
Note that this bound becomes tighter for larger $N$.

Topological defects occur, if
the Peccei-Quinn symmetry is restored after inflation,
i.e., $T_\text{R} > T_\text{PQ} \sim f_\text{PQ} = N f_a$.
This leads
to the formation of disastrous domain walls~\cite{Sikivie:1982qv}, if $N>1$, and cosmic
strings~\cite{Davis:1986xc}. The abundance of axions from cosmic strings
$\O_a^\text{str}$ exceeds that from vacuum
misalignment~\cite{Wantz:2009it,Hiramatsu:2010yu}.
Since here the reheating temperature is fixed at about $2 \D \times 10^9 \GeV$,
the situation is particularly interesting.
There are two possible cases: i) topological defects are not
created after inflation, so we do not have to care about them
or ii) they occur and we have to take them into consideration.
To avoid topological defects completely,
\begin{equation}
\label{lowerfabound}
 f_a > 2 \times 10^9 \GeV \, \frac{\Delta}{N} \,,
\end{equation}
which could lead, depending on $N$, to a stronger lower bound on $f_a$
than given above. In the case without entropy production ($\D=1$), this
favours models with $N\geq4$ such as the DFSZ model.
If~\eqref{lowerfabound} is violated, $N=1$---fulfilled by the KSVZ
model---still avoids domain walls 
and the axion density from strings,
$\O_a^\text{str} \sim 10 \times \O_a^\text{mis}$~\cite{Wantz:2009it,Hiramatsu:2010yu},
gives a tighter upper bound than~\eqref{upperfaboundstan},
\begin{equation} \label{upperfaboundstrings}
 f_a < 1.3 \times 10^9 \GeV \quad \text{for} \quad
a_0^\prime = N = \Delta =1 \;,\; r=0.04 \,,
\end{equation}
where $a_0^\prime$ comprises model-dependent factors at the production
of axions from cosmic strings.
Combining the above considerations, only the small interval
$(1.3 \text{--} 2) \times 10^9\GeV$ for $f_a$ might be excluded.
In this sense, the allowed band for $f_a$ is not changed.

With entropy production ($\D>1$) the reheating temperature
is raised to compensate the dilution of the baryon asymmetry, so the
Peccei-Quinn symmetry becomes restored for a larger range of
values of $f_a$. 
We consider entropy production after the QCD phase transition
by late particle decay and estimate the axion abundance from
cosmic strings as $\O_a^\text{str} \sim 10 \times \O_a^\text{mis}/\Delta$
with $\O_a^\text{mis}$ as in~\eqref{eq:oa1}.
Then the upper bound \eqref{upperfaboundstrings} on $f_a$ is softened by
a factor of $\D^{6/7}$, because the axions are diluted, while the amount
produced remains the same.
The maximal
 $\D^\text{max}\sim 10^4$ corresponds to $f_a \lesssim 4 \times 10^{12} \GeV$,
 so in this situation the Peccei-Quinn symmetry is probably
 restored, because $T_\text{R} \sim 2 \times 10^{13} \GeV$.
We therefore list this upper bound on $f_a$ in Table~\ref{tab:1}.
Larger values are possible, if the symmetry is not restored and if one
accepts the axino to be heavier than the gluino.

 Also the upper bound on $f_a$ from vacuum misalignment
 is affected by $\D>1$ if the entropy is produced
 after
 the QCD phase transition.
In this case the Universe is likely to be dominated by matter at 
the onset of axion oscillations at $T_a^\text{osc}\sim 1 \GeV$
and the upper bound on $f_a$ becomes~\cite{Kawasaki:1995vt}
\begin{equation}
\label{eq:oabound}
 \left(\frac{f_a}{10^{14} \GeV}\right)^2 \lesssim
 \left(\frac{r}{0.02}\right) \left(\frac{4 \MeV}{T^\text{dec}}\right) \, ,
\end{equation}
where $T^\text{dec}$ denotes as before the temperature of the Universe after the
entropy-producing particle decayed.
In this case the axion decay constant is constrained more strongly by
$\O_a^\text{str}$ and too late axino and saxion decays.
Altogether, a large dilution factor opens up\hspace{0pt}---at least for $N=1$---more
 parameter space, but the situation depends on the time of entropy production.
 
We would like to point out the appealing possibility that the saxion decay
gives rise to the maximal dilution after the QCD phase transition.
For the harmless value $f_a =10^{10}\GeV$ this is the case for
an initial amplitude of saxion oscillations equal to the geometric mean
of the two scales involved, i.e.,
$\phi^\text{i}_\text{sax} \sim \sqrt{f_a \mplanck}$, and a quite
small saxion mass $\msax \sim 10 \GeV$~\cite{Hasenkamp:2010if}.
However, it is understood that 
an adjusted initial amplitude might produce the desired amount of entropy
 for any of the allowed parameter values.
  
The given bounds from the different production mechanisms
of axions refer in each case to ``standard values'' in
parameter space. They can be relaxed or circumvented
in ``non-standard'' scenarios. Constructing models that realise
a small $\O_a$ is beyond the scope of this letter.

\section{Conclusions and outlook}
We have presented how
the constraints on the axion multiplet
in scenarios of thermal leptogenesis with
gravitino dark matter
are relaxed, if R-parity is broken.
Our results are comprised in Table~\ref{tab:1}.
Obviously, the most important findings also hold in less restricted
scenarios.

The saxion mass becomes practically
unconstrained. 
An axino NLSP stays excluded, but
the axino may be anywhere else
in the superparticle mass spectrum as long as
a decay channel into an
ordinary superparticle of the MSSM
is kinematically open and allowed at
tree-level in the low-energy effective
theory.
 This does not
hold if the only possible decay 
is into a light fermion (up-quark,
electron, neutrino) and its
superpartner. 
In turn, for the DFSZ axion model, usually
considered superparticle spectra
are possible, 
\begin{table}[ht]
 \label{tab:1}
\centering
  \begin{tabular}{|c|c|c|c|}
  \hline
 [GeV]&	standard	&	$\slashed{R}$		&	$\slashed{R} \land \D|_{T<\L_\text{QCD}}=\D^\text{max}$
\\ \hline
$f_a$	& $\mathbf{\lesssim 10^{10}}$ & $\lesssim 10^{10} $ &  $\lesssim 4 \times 10^{12}  $ 
\\ \hline
$\maxino$	& $\begin{array}{c} > \mgluino \\ ${\footnotesize (or $>m_{\widetilde H}+m_h$ in DFSZ)}$\end{array}$	& $ > m_\text{``nlsp''}$ & 
$ \begin{array}{c} \gtrsim \text{max}[ m_\text{``nlsp''} , \qquad \\ \qquad 300  \left(\frac{f_a}{10^{11} \GeV}\right)^\frac{2}{3}] \end{array}$
\\ \hline
$m_\text{sax}$	& $ \begin{array}{c}  > 760 \left(\frac{f_a}{10^{10} \GeV}\right)^\frac{2}{3}\\$
{\footnotesize (or $>2 m_h$ in DFSZ)}
 $ \\ \text{or }
\in [  5  \left(\frac{f_a}{10^{10} \GeV}\right)^\frac{2}{3}  ,2 m_\text{nlsp} ] \end{array}$  & $\gtrsim 5  \left(\frac{f_a}{10^{10} \GeV}\right)^\frac{2}{3}$ 
 & $\gtrsim 5 \left(\frac{f_a}{10^{10} \GeV}\right)^\frac{2}{3}$ \\ \hline 
 $\phi^\text{i}_\text{sax}$ & $\lesssim  10^{13} \text{--} 10^{15}$ & $\lesssim 10^{13} \text{--} 10^{15}$ & $\lesssim 5 \times (10^{14} \text{--}  10^{16})$ \\ \hline
\end{tabular}

\caption{Parameter constraints for the different scenarios 
(standard: R-parity conserved ($\D=1$), 
$\slashed{R}$: R-parity violated ($\D=1$) and
violated R-parity with the maximal entropy dilution at the right time). 
Units are GeV where not written explicitly.
 Here, we assume the self-coupling $x \ll 1$.
 Only if a mass bound
is sensitive to the actual value of $f_a$, its dependence is given. 
By ``nlsp'' we indicate that the bound actually does not hold for any possible NLSP and depends
 on the axion model.
 The upper bound on $f_a$ in the standard
scenario is boldface to indicate that it arises from $\O_a$ and the (KSVZ) axino decay. In the
$\slashed{R}$ case it stems from $\O_a^\text{mis}$ only. If, furthermore, matter dominates at $T^\text{osc}_a$
it stems from $\O_a^\text{str}$. 
}
\end{table}
including spectra with stau NLSP\@.
The sufficient condition is that the axino
decay into the NLSP is not too strongly
kinematically suppressed. 
In the KSVZ model, decays into fermion-sfermion pairs are loop-suppressed.
However, in both models
an open axino decay channel into a neutralino
suffices.
As a consequence, it will be interesting to construct concrete models with
the naturally expected axino and saxion mass of order $m_\text{susy}$,
possibly along the lines of existing models like~\cite{Jeong:2011xu,Carpenter:2009sw,Choi:2011xt}.

Constraints on the axion decay constant from axino and saxion decay
are softened. Interestingly, the DFSZ model is---depending on the $\mu$-parameter---less restricted
from particle decay already in the standard scenario. For example,
the saxion mass could be unconstrained, if the Higgs boson is lighter
than the NLSP\@.
Since in the considered setting the reheating temperature is fixed at a
large value, the constraints from axion overproduction on the
decay constant 
are particularly interesting and depend on the axion model.
However, they are not
affected by R-parity violation.
Therefore we have shown how late-time entropy production
might soften the upper bounds on the axion
decay constant and the initial amplitude of
saxion oscillations after inflation.

There is no constraint on the saxion-axion-axion self-coupling
as long as $f_a < 10^{10} \GeV$ and $\msax >100 \GeV$.
A large self-coupling of order one can remove constraints on
the saxion mass, especially in the standard scenario.
We remark that for suitable values of the self-coupling and other model
parameters one can obtain any desired amount of additional radiation
energy in the Universe formed by axions.
It will be interesting to identify (more) models
realising a particular self-coupling
and/or a small axion density $\O_a\ll\O_\text{DM}$.

Furthermore, it might be interesting to investigate
if other scenarios of dark and visible matter
enable a solution of the strong CP problem.
We conclude that broken R-parity does not
only solve the NLSP decay problem but
also makes it easier to solve the strong CP problem.
Our results are an additional motivation for
scenarios of thermal leptogenesis with gravitino 
dark matter and broken R-parity.

\subsection*{Acknowledgements}
We would like to thank Georg Raffelt, Wilfried Buchm\"uller, Laura Covi and  Kazunori Nakayama  for
helpful discussions.
J.K.\ thanks CINVESTAV in Mexico City for hospitality during stages
of this work.
This work was supported by the German Science Foundation (DFG) via the
Junior Research Group ``SUSY Phenomenology'' within the Collaborative
Research Centre 676 ``Particles, Strings and the Early Universe''.

\phantomsection 
\addcontentsline{toc}{chapter}{References}
\bibliography{ARbibliography}

\providecommand{\href}[2]{#2}\begingroup\raggedright\begin{thebibliography}{10}

\bibitem{Fukugita:1986hr}
M.~Fukugita and T.~Yanagida, ``{Baryogenesis Without Grand Unification}'',
\href{http://dx.doi.org/10.1016/0370-2693(86)91126-3}{{\em Phys. Lett.} {\bf
  B174} (1986)  45}.

\bibitem{Falomkin:1984eu}
I.~V. Falomkin {\em et al.}, ``Low-energy anti-p {H}e-4 annihilation and
  problems of the modern cosmology, {GUT} and {SUSY} models'', {\em Nuovo Cim.}
  {\bf A79} (1984)  193--204.
[Yad.\ Fiz.\ {\bf 39} (1984) 990].

\bibitem{Ellis:1984eq}
J.~R. Ellis, J.~E. Kim, and D.~V. Nanopoulos, ``{Cosmological Gravitino
  Regeneration and Decay}'',
\href{http://dx.doi.org/10.1016/0370-2693(84)90334-4}{{\em Phys. Lett.} {\bf
  B145} (1984)  181}.

\bibitem{Khlopov:1984pf}
M.~Y. Khlopov and A.~D. Linde, ``{Is It Easy to Save the Gravitino?}'',
\href{http://dx.doi.org/10.1016/0370-2693(84)91656-3}{{\em Phys. Lett.} {\bf
  B138} (1984)  265--268}.

\bibitem{Peccei:1977hh}
R.~D. Peccei and H.~R. Quinn, ``{CP Conservation in the Presence of
  Instantons}'',
\href{http://dx.doi.org/10.1103/PhysRevLett.38.1440}{{\em Phys. Rev. Lett.}
  {\bf 38} (1977)  1440--1443}.

\bibitem{Peccei:1977ur}
R.~D. Peccei and H.~R. Quinn, ``{Constraints Imposed by CP Conservation in the
  Presence of Instantons}'',
\href{http://dx.doi.org/10.1103/PhysRevD.16.1791}{{\em Phys. Rev.} {\bf D16}
  (1977)  1791--1797}.

\bibitem{Moroi:1993mb}
T.~Moroi, H.~Murayama, and M.~Yamaguchi, ``Cosmological constraints on the
  light stable gravitino'',
\href{http://dx.doi.org/10.1016/0370-2693(93)91434-O}{{\em Phys. Lett.} {\bf
  B303} (1993)  289--294}.

\bibitem{Hasenkamp:2010if}
J.~Hasenkamp and J.~Kersten, ``{Leptogenesis, Gravitino Dark Matter and Entropy
  Production}'', \href{http://dx.doi.org/10.1103/PhysRevD.82.115029}{{\em
  Phys.Rev.} {\bf D82} (2010)  115029},
\href{http://arxiv.org/abs/1008.1740}{{\tt arXiv:1008.1740 [hep-ph]}}.

\bibitem{Buchmuller:2007ui}
W.~Buchm{\"u}ller, L.~Covi, K.~Hamaguchi, A.~Ibarra, and T.~Yanagida,
  ``{Gravitino dark matter in R-parity breaking vacua}'',
  \href{http://dx.doi.org/10.1088/1126-6708/2007/03/037}{{\em JHEP} {\bf 03}
  (2007)  037},
\href{http://arxiv.org/abs/hep-ph/0702184}{{\tt arXiv:hep-ph/0702184}}.

\bibitem{Dine:1981rt}
M.~Dine, W.~Fischler, and M.~Srednicki, ``{A Simple Solution to the Strong CP
  Problem with a Harmless Axion}'',
\href{http://dx.doi.org/10.1016/0370-2693(81)90590-6}{{\em Phys. Lett.} {\bf
  B104} (1981)  199}.

\bibitem{Zhitnitsky:1980tq}
A.~R. Zhitnitsky, ``{On Possible Suppression of the Axion Hadron
  Interactions}'', {\em Sov. J. Nucl. Phys.} {\bf 31} (1980)  260.
[Yad.\ Fiz.\ {\bf 31} (1980) 497].

\bibitem{Kim:1979if}
J.~E. Kim, ``{Weak Interaction Singlet and Strong CP Invariance}'',
\href{http://dx.doi.org/10.1103/PhysRevLett.43.103}{{\em Phys. Rev. Lett.} {\bf
  43} (1979)  103}.

\bibitem{Shifman:1979if}
M.~A. Shifman, A.~I. Vainshtein, and V.~I. Zakharov, ``{Can Confinement Ensure
  Natural CP Invariance of Strong Interactions?}'',
\href{http://dx.doi.org/10.1016/0550-3213(80)90209-6}{{\em Nucl. Phys.} {\bf
  B166} (1980)  493}.

\bibitem{Witten:1984dg}
E.~Witten, ``{Some Properties of O(32) Superstrings}'',
\href{http://dx.doi.org/10.1016/0370-2693(84)90422-2}{{\em Phys. Lett.} {\bf
  B149} (1984)  351--356}.

\bibitem{Covi:2001nw}
L.~Covi, H.-B. Kim, J.~E. Kim, and L.~Roszkowski, ``{Axinos as dark matter}'',
  \href{http://dx.doi.org/10.1088/1126-6708/2001/05/033}{{\em JHEP} {\bf 05}
  (2001)  033},
\href{http://arxiv.org/abs/hep-ph/0101009}{{\tt arXiv:hep-ph/0101009}}.

\bibitem{Brandenburg:2004du}
A.~Brandenburg and F.~D. Steffen, ``{Axino dark matter from thermal
  production}'', \href{http://dx.doi.org/10.1088/1475-7516/2004/08/008}{{\em
  JCAP} {\bf 0408} (2004)  008},
\href{http://arxiv.org/abs/hep-ph/0405158}{{\tt arXiv:hep-ph/0405158}}.

\bibitem{Strumia:2010aa}
A.~Strumia, ``{Thermal production of axino Dark Matter}'',
  \href{http://dx.doi.org/10.1007/JHEP06(2010)036}{{\em JHEP} {\bf 06} (2010)
  036},
\href{http://arxiv.org/abs/1003.5847}{{\tt arXiv:1003.5847 [hep-ph]}}.

\bibitem{Choi:2008zq}
K.-Y. Choi, J.~E. Kim, H.~M. Lee, and O.~Seto, ``{Neutralino dark matter from
  heavy axino decay}'',
  \href{http://dx.doi.org/10.1103/PhysRevD.77.123501}{{\em Phys. Rev.} {\bf
  D77} (2008)  123501},
\href{http://arxiv.org/abs/0801.0491}{{\tt arXiv:0801.0491 [hep-ph]}}.

\bibitem{Baer:2011hx}
H.~Baer, A.~Lessa, S.~Rajagopalan, and W.~Sreethawong, ``{Mixed
  axion/neutralino cold dark matter in supersymmetric models}'',
\href{http://arxiv.org/abs/1103.5413}{{\tt arXiv:1103.5413 [hep-ph]}}.

\bibitem{Chun:1995hc}
E.~J. Chun and A.~Lukas, ``{Axino mass in supergravity models}'',
  \href{http://dx.doi.org/10.1016/0370-2693(95)00881-K}{{\em Phys. Lett.} {\bf
  B357} (1995)  43--50},
\href{http://arxiv.org/abs/hep-ph/9503233}{{\tt arXiv:hep-ph/9503233}}.

\bibitem{Jeong:2011xu}
K.~S. Jeong and M.~Yamaguchi, ``{Axion model in gauge-mediated supersymmetry
  breaking and a solution to the $\mu/B\mu$ problem}'',
\href{http://arxiv.org/abs/1102.3301}{{\tt arXiv:1102.3301 [hep-ph]}}.

\bibitem{Baer:2010gr}
H.~Baer, S.~Kraml, A.~Lessa, and S.~Sekmen, ``{Thermal leptogenesis and the
  gravitino problem in the Asaka-Yanagida axion/axino dark matter scenario}'',
  \href{http://dx.doi.org/10.1088/1475-7516/2011/04/039}{{\em JCAP} {\bf 1104}
  (2011)  039},
\href{http://arxiv.org/abs/1012.3769}{{\tt arXiv:1012.3769 [hep-ph]}}.

\bibitem{Kawasaki:2011ym}
M.~Kawasaki, N.~Kitajima, and K.~Nakayama, ``{Cosmological Aspects of Inflation
  in a Supersymmetric Axion Model}'',
\href{http://arxiv.org/abs/1104.1262}{{\tt arXiv:1104.1262 [hep-ph]}}.

\bibitem{Kawasaki:2007mk}
M.~Kawasaki, K.~Nakayama, and M.~Senami, ``{Cosmological implications of
  supersymmetric axion models}'',
  \href{http://dx.doi.org/10.1088/1475-7516/2008/03/009}{{\em JCAP} {\bf 0803}
  (2008)  009},
\href{http://arxiv.org/abs/0711.3083}{{\tt arXiv:0711.3083 [hep-ph]}}.

\bibitem{Pospelov:2010hj}
M.~Pospelov and J.~Pradler, ``{Big Bang Nucleosynthesis as a Probe of New
  Physics}'', \href{http://dx.doi.org/10.1146/annurev.nucl.012809.104521}{{\em
  Ann. Rev. Nucl. Part. Sci.} {\bf 60} (2010)  539--568},
\href{http://arxiv.org/abs/1011.1054}{{\tt arXiv:1011.1054 [hep-ph]}}.

\bibitem{Aver:2010wq}
E.~Aver, K.~A. Olive, and E.~D. Skillman, ``{A New Approach to Systematic
  Uncertainties and Self- Consistency in Helium Abundance Determinations}'',
  \href{http://dx.doi.org/10.1088/1475-7516/2010/05/003}{{\em JCAP} {\bf 1005}
  (2010)  003},
\href{http://arxiv.org/abs/1001.5218}{{\tt arXiv:1001.5218}}.

\bibitem{Izotov:2010ca}
Y.~I. Izotov and T.~X. Thuan, ``{The primordial abundance of 4He: evidence for
  non-standard big bang nucleosynthesis}'',
  \href{http://dx.doi.org/10.1088/2041-8205/710/1/L67}{{\em Astrophys. J.} {\bf
  710} (2010)  L67--L71},
\href{http://arxiv.org/abs/1001.4440}{{\tt arXiv:1001.4440 [astro-ph.CO]}}.

\bibitem{Dunkley:2010ge}
J.~Dunkley {\em et al.}, ``{The Atacama Cosmology Telescope: Cosmological
  Parameters from the 2008 Power Spectra}'',
\href{http://arxiv.org/abs/1009.0866}{{\tt arXiv:1009.0866 [astro-ph.CO]}}.

\bibitem{Hamann:2007pi}
J.~Hamann, S.~Hannestad, G.~G. Raffelt, and Y.~Y.~Y. Wong, ``{Observational
  bounds on the cosmic radiation density}'',
  \href{http://dx.doi.org/10.1088/1475-7516/2007/08/021}{{\em JCAP} {\bf 0708}
  (2007)  021},
\href{http://arxiv.org/abs/0705.0440}{{\tt arXiv:0705.0440 [astro-ph]}}.

\bibitem{Kawasaki:1999na}
M.~Kawasaki, K.~Kohri, and N.~Sugiyama, ``{Cosmological Constraints on
  Late-time Entropy Production}'',
  \href{http://dx.doi.org/10.1103/PhysRevLett.82.4168}{{\em Phys. Rev. Lett.}
  {\bf 82} (1999)  4168},
\href{http://arxiv.org/abs/astro-ph/9811437}{{\tt arXiv:astro-ph/9811437}}.

\bibitem{Adhya:2003tr}
P.~Adhya, D.~R. Chaudhuri, and S.~Hannestad, ``{Late-time Entropy Production
  from Scalar Decay and Relic Neutrino Temperature}'',
  \href{http://dx.doi.org/10.1103/PhysRevD.68.083519}{{\em Phys. Rev.} {\bf
  D68} (2003)  083519},
\href{http://arxiv.org/abs/astro-ph/0309135}{{\tt arXiv:astro-ph/0309135}}.

\bibitem{Hannestad:2004px}
S.~Hannestad, ``{What is the lowest possible reheating temperature?}'',
  \href{http://dx.doi.org/10.1103/PhysRevD.70.043506}{{\em Phys. Rev.} {\bf
  D70} (2004)  043506},
\href{http://arxiv.org/abs/astro-ph/0403291}{{\tt arXiv:astro-ph/0403291}}.

\bibitem{Kawasaki:2000en}
M.~Kawasaki, K.~Kohri, and N.~Sugiyama, ``{MeV-scale reheating temperature and
  thermalization of neutrino background}'',
  \href{http://dx.doi.org/10.1103/PhysRevD.62.023506}{{\em Phys. Rev.} {\bf
  D62} (2000)  023506},
\href{http://arxiv.org/abs/astro-ph/0002127}{{\tt arXiv:astro-ph/0002127}}.

\bibitem{Ichikawa:2005vw}
K.~Ichikawa, M.~Kawasaki, and F.~Takahashi, ``{The oscillation effects on
  thermalization of the neutrinos in the universe with low reheating
  temperature}'', \href{http://dx.doi.org/10.1103/PhysRevD.72.043522}{{\em
  Phys. Rev.} {\bf D72} (2005)  043522},
\href{http://arxiv.org/abs/astro-ph/0505395}{{\tt arXiv:astro-ph/0505395}}.

\bibitem{Farrar:2010ps}
G.~R. Farrar, R.~Mackeprang, D.~Milstead, and J.~P. Roberts, ``{Limit on the
  mass of a long-lived or stable gluino}'',
  \href{http://dx.doi.org/10.1007/JHEP02(2011)018}{{\em JHEP} {\bf 02} (2011)
  018},
\href{http://arxiv.org/abs/1011.2964}{{\tt arXiv:1011.2964 [hep-ph]}}.

\bibitem{Kim:1998va}
J.~E. Kim, ``{Constraints on Very Light Axions from Cavity Experiments}'',
  \href{http://dx.doi.org/10.1103/PhysRevD.58.055006}{{\em Phys. Rev.} {\bf
  D58} (1998)  055006},
\href{http://arxiv.org/abs/hep-ph/9802220}{{\tt arXiv:hep-ph/9802220}}.

\bibitem{Nieves:1986ed}
J.~F. Nieves, ``{Radiative photino decay in models with an invisible axion}'',
\href{http://dx.doi.org/10.1016/0370-2693(86)91026-9}{{\em Phys. Lett.} {\bf
  B174} (1986)  411}.

\bibitem{Aaltonen:2009kea}
CDF, T.~Aaltonen {\em et al.}, ``{Search for Long-Lived Massive Charged
  Particles in 1.96 TeV $\bar{p}p$ Collisions}'',
  \href{http://dx.doi.org/10.1103/PhysRevLett.103.021802}{{\em Phys. Rev.
  Lett.} {\bf 103} (2009)  021802},
\href{http://arxiv.org/abs/0902.1266}{{\tt arXiv:0902.1266 [hep-ex]}}.

\bibitem{Covi:2002vw}
L.~Covi, L.~Roszkowski, and M.~Small, ``{Effects of squark processes on the
  axino CDM abundance}'',
  \href{http://dx.doi.org/10.1088/1126-6708/2002/07/023}{{\em JHEP} {\bf 07}
  (2002)  023},
\href{http://arxiv.org/abs/hep-ph/0206119}{{\tt arXiv:hep-ph/0206119}}.

\bibitem{Covi:2004rb}
L.~Covi, L.~Roszkowski, R.~Ruiz~de Austri, and M.~Small, ``{Axino dark matter
  and the CMSSM}'', \href{http://dx.doi.org/10.1088/1126-6708/2004/06/003}{{\em
  JHEP} {\bf 06} (2004)  003},
\href{http://arxiv.org/abs/hep-ph/0402240}{{\tt arXiv:hep-ph/0402240}}.

\bibitem{Hooper:2004qf}
D.~Hooper and L.-T. Wang, ``{Evidence for axino dark matter in the galactic
  bulge}'', \href{http://dx.doi.org/10.1103/PhysRevD.70.063506}{{\em Phys.
  Rev.} {\bf D70} (2004)  063506},
\href{http://arxiv.org/abs/hep-ph/0402220}{{\tt arXiv:hep-ph/0402220}}.

\bibitem{Chun:2006ss}
E.~J. Chun and H.~B. Kim, ``{Axino Light Dark Matter and Neutrino Masses with
  R-parity Violation}'',
  \href{http://dx.doi.org/10.1088/1126-6708/2006/10/082}{{\em JHEP} {\bf 10}
  (2006)  082},
\href{http://arxiv.org/abs/hep-ph/0607076}{{\tt arXiv:hep-ph/0607076}}.

\bibitem{Kim:2001sh}
H.-B. Kim and J.~E. Kim, ``{Late decaying axino as CDM and its lifetime
  bound}'', \href{http://dx.doi.org/10.1016/S0370-2693(01)01507-6}{{\em Phys.
  Lett.} {\bf B527} (2002)  18--22},
\href{http://arxiv.org/abs/hep-ph/0108101}{{\tt arXiv:hep-ph/0108101}}.

\bibitem{Lyth:1993zw}
D.~H. Lyth, ``{Dilution of cosmological densities by saxino decay}'',
  \href{http://dx.doi.org/10.1103/PhysRevD.48.4523}{{\em Phys. Rev.} {\bf D48}
  (1993)  4523--4533},
\href{http://arxiv.org/abs/hep-ph/9306293}{{\tt arXiv:hep-ph/9306293}}.

\bibitem{Davidson:2002qv}
S.~Davidson and A.~Ibarra, ``{A lower bound on the right-handed neutrino mass
  from leptogenesis}'',
  \href{http://dx.doi.org/10.1016/S0370-2693(02)01735-5}{{\em Phys. Lett.} {\bf
  B535} (2002)  25--32},
\href{http://arxiv.org/abs/hep-ph/0202239}{{\tt arXiv:hep-ph/0202239}}.

\bibitem{Raffelt:2006cw}
G.~G. Raffelt, ``{Astrophysical axion bounds}'',
  \href{http://dx.doi.org/10.1007/978-3-540-73518-2_3}{{\em Lect. Notes Phys.}
  {\bf 741} (2008)  51--71},
\href{http://arxiv.org/abs/hep-ph/0611350}{{\tt arXiv:hep-ph/0611350}}.

\bibitem{Preskill:1982cy}
J.~Preskill, M.~B. Wise, and F.~Wilczek, ``{Cosmology of the invisible
  axion}'',
\href{http://dx.doi.org/10.1016/0370-2693(83)90637-8}{{\em Phys. Lett.} {\bf
  B120} (1983)  127--132}.

\bibitem{Abbott:1982af}
L.~F. Abbott and P.~Sikivie, ``{A cosmological bound on the invisible axion}'',
\href{http://dx.doi.org/10.1016/0370-2693(83)90638-X}{{\em Phys. Lett.} {\bf
  B120} (1983)  133--136}.

\bibitem{Dine:1982ah}
M.~Dine and W.~Fischler, ``{The not-so-harmless axion}'',
\href{http://dx.doi.org/10.1016/0370-2693(83)90639-1}{{\em Phys. Lett.} {\bf
  B120} (1983)  137--141}.

\bibitem{Sikivie:1982qv}
P.~Sikivie, ``{Of Axions, Domain Walls and the Early Universe}'',
\href{http://dx.doi.org/10.1103/PhysRevLett.48.1156}{{\em Phys. Rev. Lett.}
  {\bf 48} (1982)  1156--1159}.

\bibitem{Davis:1986xc}
R.~L. Davis, ``{Cosmic Axions from Cosmic Strings}'',
\href{http://dx.doi.org/10.1016/0370-2693(86)90300-X}{{\em Phys. Lett.} {\bf
  B180} (1986)  225}.

\bibitem{Wantz:2009it}
O.~Wantz and E.~P.~S. Shellard, ``{Axion Cosmology Revisited}'',
  \href{http://dx.doi.org/10.1103/PhysRevD.82.123508}{{\em Phys. Rev.} {\bf
  D82} (2010)  123508},
\href{http://arxiv.org/abs/0910.1066}{{\tt arXiv:0910.1066 [astro-ph.CO]}}.

\bibitem{Hiramatsu:2010yu}
T.~Hiramatsu, M.~Kawasaki, T.~Sekiguchi, M.~Yamaguchi, and J.~Yokoyama,
  ``{Improved estimation of radiated axions from cosmological axionic
  strings}'',
\href{http://arxiv.org/abs/1012.5502}{{\tt arXiv:1012.5502}}.

\bibitem{Kawasaki:1995vt}
M.~Kawasaki, T.~Moroi, and T.~Yanagida, ``{Can decaying particles raise the
  upperbound on the Peccei-Quinn scale?}'',
  \href{http://dx.doi.org/10.1016/0370-2693(96)00743-5}{{\em Phys. Lett.} {\bf
  B383} (1996)  313--316},
\href{http://arxiv.org/abs/hep-ph/9510461}{{\tt arXiv:hep-ph/9510461}}.

\bibitem{Carpenter:2009sw}
L.~M. Carpenter, M.~Dine, G.~Festuccia, and L.~Ubaldi, ``{Axions in Gauge
  Mediation}'', \href{http://dx.doi.org/10.1103/PhysRevD.80.125023}{{\em Phys.
  Rev.} {\bf D80} (2009)  125023},
\href{http://arxiv.org/abs/0906.5015}{{\tt arXiv:0906.5015 [hep-th]}}.

\bibitem{Choi:2011xt}
K.~Choi, K.~S. Jeong, K.-I. Okumura, and M.~Yamaguchi, ``{Mixed Mediation of
  Supersymmetry Breaking with Anomalous U(1) Gauge Symmetry}'',
\href{http://arxiv.org/abs/1104.3274}{{\tt arXiv:1104.3274 [hep-ph]}}.

\end{thebibliography}\endgroup

\end{document}